\documentclass[pdflatex,iicol,sn-mathphys-num]{sn-jnl}

\usepackage{amssymb}
\usepackage{bm}
\usepackage{mathrsfs}
\usepackage{mathtools}

\let\a=\alpha
\let\m=\mu
\let\p=\phi
\let\t=\theta
\let\de=\partial
\DeclareMathOperator{\am}{am}
\DeclareMathOperator{\cd}{cd}
\DeclareMathOperator{\sn}{sn}
\DeclareMathOperator{\dn}{dn}
\newcommand{\Ea}{\mathscr{E}}
\newcommand{\Ha}{\mathscr{H}}
\newcommand{\La}{\mathscr{L}}
\newcommand{\dd}{\text{d}}
\newcommand{\vek}[1]{\bm{#1}}
\newcommand{\grad}{\vek\nabla}
\newcommand{\divg}{\grad\cdot}

\begin{document}

\title{Chiral soliton lattice in inhomogeneous magnetic fields}

\author[1]{\fnm{Tom\'a\v{s}} \sur{Brauner}}

\author[2]{\fnm{Ramkumar} \sur{Radhakrishnan}}

\affil[1]{\orgdiv{Department of Mathematics and Physics}, \orgname{University of Stavanger}, \orgaddress{\city{Stavanger}, \postcode{N-4036}, \country{Norway}}}

\affil[2]{\orgdiv{Department of Physics and Astronomy}, \orgname{North Carolina State University},\\ \orgaddress{\city{Raleigh}, \postcode{NC 27695}, \country{USA}}}

\abstract{It has been known that in sufficiently strong uniform magnetic fields, the ground state of quantum chromodynamics (QCD) supports a spatially modulated condensate of neutral pions, dubbed chiral soliton lattice (CSL). In this paper, we investigate whether a similar ordered ground state might exist when the external magnetic field is nonuniform, as appropriate for potential phenomenological applications, including heavy-ion collisions and neutron stars. To that end, we use the low-energy effective field theory of QCD, restricted to the neutral pions as the sole low-energy degrees of freedom in strong magnetic fields. In the limit of vanishing pion mass, we achieve a complete characterization of magnetic fields supporting a CSL-like ground state. Moreover, for a simple but infinite family of magnetic fields, we find the corresponding CSL state analytically. Going away from the massless limit requires full numerical minimization of the energy functional. Here we provide some sample numerical results, focusing on the qualitative differences as compared to the situations with a uniform magnetic field or a vanishing pion mass. The main conclusion remains unchanged: while bending the magnetic field typically reduces the energy gain due to the neutral pion condensation, a CSL-type ground state is still possible. As a byproduct of our work, we map the location of the CSL phase in the phase diagram of QCD in a uniform magnetic field and finite volume.}

\keywords{Chiral anomaly, chiral soliton lattice, effective field theory, phase diagram of QCD.}

\maketitle


\section{Introduction}
\label{sec:intro}

The phase diagram of quantum chromodynamics (QCD) in the presence of a strong magnetic field continues to attract considerable attention due to its relevance for the physics of heavy-ion collisions and neutron stars~\cite{Andersen:2014xxa,Miransky:2015ava}. It has been known for a long time that the magnetic field induces a 't Hooft anomaly that significantly affects the low-energy physics of QCD. More recently, it was discovered that the magnetic field can also have a direct impact on the phase structure of QCD through the formation of a condensate of neutral pions; see Refs.~\cite{Son2004a,Son:2007ny,Bergman:2008qv,Thompson:2008qw,Rebhan:2008ur,Eto2013a} for some early works on the subject. This condensate is spatially nonuniform and by virtue of the chiral anomaly carries a nonzero baryon number. The precise dependence of the ground state of QCD with two quark flavors on the baryon chemical potential and the external magnetic field was mapped using a model-independent low-energy effective field theory (EFT) in Ref.~\cite{Brauner:2016pko}. Owing to the spatial periodicity and the topological nature of the pion condensate, the new state of matter was dubbed chiral soliton lattice (CSL) in analogy with similar phenomena observed in certain types of magnets~\cite{Togawa2012a,Kishine2015a}. Simultaneously, it was found that the QCD vacuum and the neutral-pion CSL are not the only relevant candidate ground states. In sufficiently strong magnetic fields, it may be energetically advantageous to form a secondary condensate of charged pions.

These findings helped inspire a new line of research, with numerous subsequent works investigating similar anomaly-induced states for various combinations of physical theories of quark matter and external conditions. In a relatively quick succession, a CSL-type ground state was found in QCD under rotation and at nonzero baryon (and possibly isospin) chemical potential~\cite{Huang:2017pqe,Nishimura:2020odq,Eto:2021gyy,Eto:2023rzd}, QCD-like theories featuring a different gauge group or a representation thereof than QCD itself, exposed to an external magnetic field~\cite{Brauner:2019rjg,Brauner:2019aid}, QCD in the presence of a magnetic field and isospin chemical potential~\cite{Gronli:2022cri}, and QCD exposed to an oscillating laser field~\cite{Yamada:2021jhy}.

In parallel, further investigations were made into the structure of the excitation spectrum above the CSL ground state~\cite{Yamamoto:2015maz,Ozaki:2016vwu,Brauner:2017mui} and the effect of excitations on the phase diagram through thermal and quantum fluctuations~\cite{Brauner:2021sci,Brauner:2023ort}. Also, much effort was put into the exploration of the competition between the CSL state and other candidate ground states of dense quark and nuclear matter, including different proposals for the accompanying condensate of charged pions~\cite{Evans:2022hwr,Eto:2025fkt,Eto:2023wul,Evans:2023hms}, a mixture of neutral pion and $\eta$-meson condensates~\cite{Qiu:2023guy}, and crystalline baryon matter~\cite{Chen:2021vou,Qiu:2024zpg,Amari:2025twm}. Last but not least, the exploration of the dynamical formation of CSL-like states was initiated in Refs.~\cite{Eto:2022lhu,Higaki:2022gnw,Nishimura:2023czx,Eto:2025ebz}.

In this paper, we address an aspect of the CSL that seems to have been overlooked so far. Namely, in those real physical systems for which the CSL phase might possibly be phenomenologically relevant, the magnetic field is far from uniform. This is certainly the case for heavy-ion collisions, where the magnetic field in addition rapidly changes with time. However, even the magnetic fields of neutron stars are inhomogeneous, albeit arguably at distances much longer than the characteristic length scale of QCD. Nevertheless, at this stage, our ambition is not to make directly testable predictions. Rather, we wish to gain basic insight into how the inhomogeneous nature of the magnetic field affects the CSL solution. Do nonuniform magnetic fields tend to destroy the CSL order, and if yes, in what kind of fields does it survive? Or is it possible that a suitably engineered nonuniform magnetic field might in fact further stabilize the CSL state?

Owing to the exploratory nature of these questions, we adopt the simplest mathematical framework possible. We only consider the two lightest quark flavors, assuming equal masses, coupled to nonzero baryon chemical potential. We use the low-energy EFT of two-flavor QCD, and simplify it further by dropping the charged pion degrees of freedom, which become heavy and thus decouple from low-energy physics in a large range of magnetic fields on account of Landau level quantization. Thus, we only keep the neutral pion field, which essentially reduces the EFT to the sine-Gordon theory, augmented with an anomaly-induced term through which the magnetic field and baryon chemical potential enter. We analyze the thermodynamics of this simplified EFT in the classical approximation, in which the task to find the ground state amounts to variational minimization of the classical energy functional.

The rest of the paper is organized as follows. In Sec.~\ref{sec:uniformB}, we revisit the physics of neutral pions in a uniform magnetic field. We review the CSL basics, following closely Ref.~\cite{Brauner:2016pko}, and improve upon the original results by considering explicitly the effects of finite volume on the ground state. (These effects were studied first in the master thesis~\cite{Kar2022}.) Placing the system in a finite volume is mandatory for nonuniform magnetic fields, for which a thermodynamic limit may not even make sense.  The physics of CSL in nonuniform magnetic fields is studied in the subsequent sections. First, in Sec.~\ref{sec:nonuniformBzeroM}, we consider the much simpler case of vanishing quark mass, where a complete characterization of the CSL ground state is possible and many results can be derived analytically. The most challenging case of two-flavor QCD with massive quarks subject to inhomogeneous magnetic fields is relegated to Sec.~\ref{sec:nonuniformBnonzeroM}. This case requires full numerical minimization of the energy functional, and we use the previous analytical results to benchmark our numerical code. Finally, in Sec.~\ref{sec:summary}, we summarize and discuss our results.


\section{Chiral soliton lattice in uniform magnetic fields}
\label{sec:uniformB}

The low-energy physics of QCD with two quark flavors is described by the chiral perturbation theory, an EFT whose form is controlled by the spontaneously broken chiral symmetry of QCD~\cite{Scherer2012a}. The dynamical degrees of freedom of this EFT are the three pions, treated as elementary (pseudo)scalar fields. However, as already stressed in the introduction, the presence of a magnetic field profoundly changes the low-energy spectrum by reorganizing the charged pion states into Landau levels. As a consequence, there is a broad range of magnetic fields where the charged pions decouple, and the low-energy physics reduces to that of neutral pions alone. At the leading order within a power-counting scheme tailored to the study of the CSL phase~\cite{Brauner:2021sci}, the corresponding reduced effective Lagrangian takes the form
\begin{equation}
\La=\frac{f_\pi^2}{2}(\de\p)^2+f_\pi^2m_\pi^2(\cos\p-1)+\frac\m{4\pi^2}\vek B\cdot\grad\p.
\label{eq:efflag}
\end{equation}
Here $\p$ is a dimensionless neutral pion field, $f_\pi$ the pion decay constant, $m_\pi$ the pion mass, $\m$ the baryon chemical potential, and $\vek B$ the external magnetic field. The last term in the Lagrangian~\eqref{eq:efflag} captures the effects of the chiral anomaly on the neutral pions. A constant has been added to the second, potential part of the Lagrangian in order to set the energy density of the trivial QCD vacuum, $\p_0=0$, to zero.

Thanks to the fact that time derivatives enter the Lagrangian through a simple square, the classical ground state of our EFT will certainly be time-independent. It is therefore sufficient to restrict from now on to static (time-independent) field configurations, for which the corresponding Hamiltonian density reads
\begin{equation}
\Ha=f_\pi^2\left[\frac12(\grad\p)^2+m_\pi^2(1-\cos\p)-\vek H\cdot\grad\p\right],
\label{eq:effham}
\end{equation}
with the shorthand notation
\begin{equation}
\vek H\equiv\frac{\m\vek B}{4\pi^2f_\pi^2}.
\label{eq:defH}
\end{equation}
In most of the paper, we will assume that the baryon chemical potential $\m$ itself remains uniform (constant), which makes $\vek H$ a solenoidal field just like the magnetic field $\vek B$ itself. To avoid language clutter, we will thus frequently refer to $\vek H$ simply as a magnetic field.

When the magnetic field is uniform, we can orient the Cartesian coordinate axes so that it points along one of them. We will call the corresponding Cartesian coordinate $z$. It is clear from the form of the Hamiltonian~\eqref{eq:effham} that the classical ground state is then necessarily uniform in all directions but $z$. In the chiral limit $m_\pi\to0$ (and in the absence of explicit boundary conditions), the ground state can in fact be found immediately by a direct pointwise minimization of the Hamiltonian density, which gives
\begin{equation}
\p_0(z)=Hz=\frac{\m Bz}{4\pi^2f_\pi^2}\qquad\text{(chiral limit)},
\label{eq:vacuniformchiral}
\end{equation}
up to an additive constant, where $B\equiv|\vek B|$.

In the more physical case of massive pions, the problem can be further simplified by introducing the dimensionless coordinate $\bar z\equiv m_\pi z$.\footnote{Note that this definition differs from one used in Ref.~\cite{Brauner:2016pko}. Throughout this paper, we will use a bar to indicate dimensionless variables defined by rescaling with suitably chosen powers of $m_\pi$ and $f_\pi$.} The task to find the ground state then amounts to the minimization of a dimensionless energy functional, obtained by dividing Eq.~\eqref{eq:effham} by $f_\pi^2m_\pi^2$ and integrating over the given range of the coordinate, $\bar z\in(\bar z_\text{min},\bar z_\text{max})$,
\begin{equation}
\bar E[\p]\equiv\int_{\bar z_\text{min}}^{\bar z_\text{max}}\dd\bar z\left[\frac12(\de_{\bar z}\p)^2+1-\cos\p-\bar H\de_{\bar z}\p\right],
\label{eq:dimlessE}
\end{equation}
where $\bar{\vek H}\equiv\vek H/m_\pi$ is likewise dimensionless. Importantly, the last term in the Hamiltonian~\eqref{eq:effham} does not contribute to the variational equation of motion (EoM) thanks to $\vek H$ having a vanishing divergence. The ground state can thus be found in two steps. In the first step, one finds a general solution of the EoM. In the second step, one then determines which of the available solutions has the lowest total energy.


\subsection{Reminder: infinite-volume limit}
\label{subsec:uniformBreminder}

We start by solving the variational EoM, following closely Ref.~\cite{Brauner:2016pko}. The EoM descending from the functional~\eqref{eq:dimlessE} is $\delta\bar E/\delta \p=-\de_{\bar z}^2\p+\sin\p=0$. It has a first integral,
\begin{equation}
\frac12(\de_{\bar z}\p)^2+\cos\p=\text{const}\equiv c.
\end{equation}
Upon changing the field variable to $\t\equiv\p-\pi$, we recognize this as the energy conservation condition for a simple pendulum in dimensionless units. The general solution is then easily found using the pendulum analogy. A direct integration leads to
\begin{equation}
\t(\bar z)=2\am(\tfrac{\bar z-\bar z_0}{k},k),
\label{eq:CSLsol}
\end{equation}
where $\am$ is the Jacobi amplitude function, $k\equiv\sqrt{2/(c+1)}$ the elliptic modulus, and $\bar z_0$ a value of the coordinate where $\t=0$. Together, $k$ and $\bar z_0$ represent the two integration constants needed to fully characterize the general solution of the EoM. Inserting our solution in Eq.~\eqref{eq:dimlessE}, we find that the integrand of the energy functional can be expressed as
\begin{equation}
\bar\Ea=2-\frac{2}{k^2}+\frac{4}{k^2}\dn^2(\tfrac{\bar z-\bar z_0}{k},k)-\frac{2\bar H}{k}\dn(\tfrac{\bar z-\bar z_0}{k},k),
\label{eq:dimlessEdensity}
\end{equation}
where $\dn$ is one of Jacobi's elliptic functions.

We shall now, by way of a reminder, assume that the system is defined in the entire space. In the absence of a boundary and of boundary conditions, we need to compute the average energy density of the solution. This follows by integrating Eq.~\eqref{eq:dimlessEdensity} over one period of the solution and dividing by the period. The result is, naturally, independent of the constant shift $\bar z_0$ and only depends on the elliptic modulus $k$,
\begin{equation}
\langle\bar\Ea\rangle=4\left[\frac12\left(1-\frac1{k^2}\right)+\frac{E(k)}{k^2K(k)}-\frac{1}{kK(k)}\frac{\bar H}{\bar H_\text{CSL}}\right].
\label{eq:dimlessEdensityaverage}
\end{equation}
Here $\bar H_\text{CSL}\equiv4/\pi$, corresponding in the physical units to
\begin{equation}
B_\text{CSL}=\frac{16\pi f_\pi^2m_\pi}{\m}.
\label{eq:BCSL}
\end{equation}
Moreover, $K(k)$ and $E(k)$ are the complete elliptic integrals of the first and second kind, respectively.

The actual ground state of the system is found by minimizing the average energy density~\eqref{eq:dimlessEdensityaverage} with respect to $k\in[0,1]$ and comparing the result with the vanishing energy density of the trivial QCD vacuum. The latter turns out to prevail for $\bar H<\bar H_\text{CSL}$, or $B<B_\text{CSL}$. Above this critical magnetic field, the CSL state forms, with the elliptic modulus $k_0$ now being a function of the magnetic field and chemical potential, defined implicitly through
\begin{equation}
\frac{E(k_0)}{k_0}=\frac{\bar H}{\bar H_\text{CSL}}=\frac{\m B}{16\pi m_\pi f_\pi^2}.
\label{eq:vacuniformk}
\end{equation}
The corresponding dimensionless average energy density is
\begin{equation}
\langle\bar\Ea_0\rangle=2\left(1-\frac1{k_0^2}\right).
\label{eq:vacuniformE}
\end{equation}
These are the main results of Ref.~\cite{Brauner:2016pko}.


\subsection{Modifications due to finite volume}
\label{subsec:uniformBfiniteV}

Let us now see how the above results are modified in a finite volume. We do not impose any explicit boundary condition, and instead let the variational principle itself choose the natural boundary condition for the solution. For the sake of simplicity, we assume that the domain in which the system lives is a rectangular box such that the magnetic field $\vek B$ is aligned with one of its sides. Denoting the (dimensionless) size of the system along the magnetic field as $\bar L$, we can choose the coordinates so that the range of integration in Eq.~\eqref{eq:dimlessE} is $\bar z\in(-\bar L/2,+\bar L/2)$. Inserting the energy density~\eqref{eq:dimlessEdensity} of the CSL solution and integrating, we obtain
\begin{align}
\notag
\bar E(k,\bar z_0)={}&2L\left(1-\frac{1}{k^2}\right)+\frac4k\left[\mathcal E(\tfrac{\bar z-\bar z_0}{k},k)\right]^{+\bar L/2}_{-\bar L/2}\\
&-2\bar H\left[\am(\tfrac{\bar z-\bar z_0}{k},k)\right]^{+\bar L/2}_{-\bar L/2},
\label{eq:vacenergyaux}
\end{align}
where $\mathcal E$ is the Jacobi epsilon function.

This still depends on both integration constants, $k$ and $\bar z_0$. However, some numerical experimentation shows that the total energy is always minimized by a solution whose gradient (derivative) is an even function of $\bar z$. This leaves us with two relevant classes of solutions: those where the gradient is maximum at $\bar z=0$ and those where it is minimum at $\bar z=0$. For the first class, we can set $\bar z_0=0$, which reduces Eq.~\eqref{eq:vacenergyaux} to
\begin{equation}
\begin{split}
\bar E_\text{max}(k)={}&2\bar L\left(1-\frac1{k^2}\right)+\frac8k\mathcal E(\tfrac{\bar L}{2k},k)\\
&-4\bar H\am(\tfrac{\bar L}{2k},k).
\end{split}
\label{eq:vacenergymax}
\end{equation}
The second class of solutions can be reproduced by setting $\bar z_0=kK(k)$. The corresponding dimensionless energy as a function of $k$ can~be~cast~as
\begin{equation}
\begin{split}
\bar E_\text{min}(k)={}&2\bar L\left(1-\frac1{k^2}\right)+\frac8k\mathcal E(\tfrac{\bar L}{2k},k)\\
&-8k\sn(\tfrac{\bar L}{2k},k)\cd(\tfrac{\bar L}{2k},k)\\
&-2\bar H\bigl[\am(\tfrac{\bar L}{2k}+K(k),k)\\
&\qquad\quad+\am(\tfrac{\bar L}{2k}-K(k),k)\bigr],
\end{split}
\label{eq:vacenergymin}
\end{equation}
where $\cd$ and $\sn$ are another two of Jacobi's elliptic functions.

In a finite volume, the configuration $\p_0(\bar z)$ minimizing the energy functional~\eqref{eq:dimlessE} is nonuniform for any nonzero value of the magnetic field $\bar H$. This is because in the absence of an explicit boundary condition, the variational principle for the total energy leads to the natural boundary condition
\begin{equation}
\vek n\cdot\grad\p_0=\vek n\cdot\vek H,
\label{eq:naturalBC}
\end{equation}
where $\vek n$ is the normal vector to the boundary. In the special case of the effectively one-dimensional system~\eqref{eq:dimlessE} subject to a uniform magnetic field $\bar H$, this reduces to $\de_ z\p_0=H$ at the boundaries of the integration range. 

This means that we have to be careful when mapping the phase diagram of the CSL in a finite volume. For a generic choice of $\bar H$ and $\bar L$, we find a discrete though infinite set of values of $k$ for which solutions of the types~\eqref{eq:vacenergymax} and~\eqref{eq:vacenergymin} satisfy the natural boundary condition, and thus have even a chance of being the ground state. This is in stark contrast to the infinite-volume limit, where Eq.~\eqref{eq:CSLsol} with any $k$ is a stationary state of the energy functional. To pin down the actual ground state, we need to find which of the discrete values of $k$ consistent with the boundary condition gives the lowest total energy. 

\begin{figure}[t]
\centering
\includegraphics[width=\columnwidth]{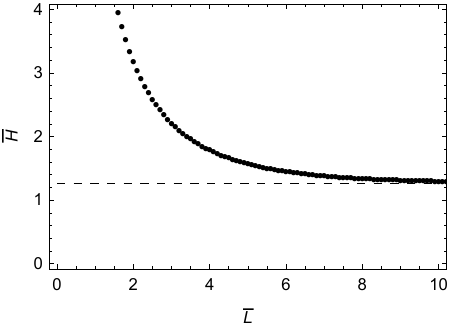}
\caption{Phase boundary between the trivial QCD vacuum and the CSL phase as a function of the dimensionless magnetic field $\bar H$ and the dimensionless system size $\bar L$. Note that for large $\bar L$, the critical magnetic field converges to $4/\pi\approx1.27$ (indicated by the horizontal dashed line), corresponding in physical units to the infinite-volume result~\eqref{eq:BCSL}.}
\label{fig:PD1d}
\end{figure}

In very weak magnetic fields, the ground state profile $\p_0(\bar z)$ will be nearly constant in the inside of the integration domain, but bend at the boundary to satisfy the natural boundary condition. In other words, the ground state will be of the type~\eqref{eq:vacenergymin}, with the gradient being minimum at the origin and increasing monotonically towards the boundary. For certain minimum value of the magnetic field, $\bar H_\text{cr}(\bar L)$, depending on the system size $\bar L$, the ground state will flip to the type~\eqref{eq:vacenergymax}, featuring a single local maximum of the gradient at the origin. This kind of solution is the finite-volume analog of the domain wall, which signals the onset of the CSL phase in infinite volume~\cite{Son:2007ny}. We therefore take such $\bar H_\text{cr}(\bar L)$ as the definition of the phase transition between the vacuum and CSL phases for finite $\bar L$. The numerical results for the critical magnetic field as a function of the system size are displayed in Fig.~\ref{fig:PD1d}.


\section{Nonuniform magnetic fields in the massless limit}
\label{sec:nonuniformBzeroM}

We now switch gears and consider a generic static magnetic field $\vek B$, defined in a finite domain $V$ in space. For the time being, we will simplify the analysis by assuming that the quarks, and hence pions, are massless. The variational problem for the energy functional obtained by integrating the Hamiltonian density~\eqref{eq:effham} over $V$ then leads to the Laplace equation for $\p$ in the bulk of $V$, accompanied by the natural, Neumann-type boundary condition~\eqref{eq:naturalBC} on the boundary $\de V$. The general theory of differential equations~\cite{Arnold2004} dictates that the Neumann boundary value problem for the Laplace equation has a unique solution up to a constant shift. We will denote the solution as $\p_0[\vek H]$. This copies to some extent what we saw in Sec.~\ref{sec:uniformB}: in the chiral limit, the energy functional has a single stationary state, which therefore automatically constitutes the absolute minimum (ground state).


\subsection{General exact results}
\label{subsec:nonuniformBzeroMgeneral}

Next, we would like to understand in greater detail the dependence of the ground state on the magnetic field, encoded in the map $\vek H\mapsto\p_0[\vek H]$ from the space of solenoidal fields to the space of harmonic functions, both in $V$. A suitable starting point is the gradient--solenoid decomposition of the magnetic field (see~\S 1.6.F of Ref.~\cite{Gallavotti2002}),
\begin{equation}
\vek H=\grad\psi[\vek H]+\vek A[\vek H],
\label{eq:helmholtz}
\end{equation}
where $\divg\vek A[\vek H]=0$ in $V$ and $\vek n\cdot\vek A[\vek H]=0$ on $\de V$. The uniqueness of this decomposition is equivalent to the uniqueness of the solution to our Neumann boundary value problem. Our ground state $\p_0[\vek H]$ corresponds exactly to the scalar potential $\psi[\vek H]$ of the magnetic field.

We will now give an alternative proof of this claim, which simultaneously shows directly that $\p_0[\vek H]=\psi[\vek H]$ up to a constant shift is the absolute minimum of the energy functional, $E[\p]$. All we need to do is complete the square under the integral,
\begin{align}
\notag
E[\p]&=f_\pi^2\int_V\left[\frac12(\grad\p)^2-(\grad\psi[\vek H]+\vek A[\vek H])\cdot\grad\p\right]\\
\notag
&=f_\pi^2\int_V\biggl[\frac12(\grad\p-\grad\psi[\vek H])^2-\frac12(\grad\psi[\vek H])^2\\
&\qquad\qquad-\vek A[\vek H]\cdot\grad\p\biggr].
\end{align}
The last term drops out upon integration thanks to the fact that $\vek A[\vek H]$ has vanishing divergence in $V$ and is tangential to the boundary $\de V$. We thus arrive at the lower bound on the total energy,
\begin{equation}
E[\p]\geq-\frac{f_\pi^2}2\int_V(\grad\psi[\vek H])^2.
\label{eq:energybound1}
\end{equation}
The lower bound is saturated if and only if $\grad\p=\grad\psi[\vek H]$ everywhere. This means that the ground state configuration $\p_0[\vek H]$ indeed equals $\psi[\vek H]$ up to an additive constant.

A simple manipulation gives a slightly weaker lower bound on the ground state energy, expressed directly in terms of the magnetic field. Namely, writing $\vek H^2=(\grad\psi[\vek H])^2+(\vek A[\vek H])^2+2\vek A[\vek H]\cdot\grad\psi[\vek H]$ and using that the integral of $\vek A[\vek H]\cdot\grad\psi[\vek H]$ over $V$ vanishes, we can extend the bound~\eqref{eq:energybound1} to
\begin{equation}
E[\p]\geq-\frac{f_\pi^2}2\int_V(\grad\psi[\vek H])^2\geq-\frac{f_\pi^2}2\int_V\vek H^2.
\label{eq:energybound2}
\end{equation}
The second inequality is saturated if and only if $\vek A[\vek H]$ vanishes everywhere in $V$.

We now have all we need for a detailed characterization of the dependence of the ground state $\p_0[\vek H]$ on the magnetic field. First, the energy of the CSL ground state, given by the right-hand side of Eq.~\eqref{eq:energybound1}, is always smaller than or equal to zero. Thus, the trivial QCD vacuum is necessarily unstable with respect to the formation of a CSL unless the $\grad\psi[\vek H]$ component of the magnetic field vanishes, which happens for the class of magnetic fields that are tangential to the boundary everywhere on $\de V$.

Second, the minimum possible energy as given by the right-hand side of Eq.~\eqref{eq:energybound2} is achieved if and only if the magnetic field is conservative (which is synonymous to the vanishing of $\vek A[\vek H]$). This guarantees that the decomposition~\eqref{eq:helmholtz} only includes the gradient part regardless of the precise choice of the domain $V$. It follows that for conservative magnetic fields, the ground state in the massless limit is insensitive to boundary effects. This is ultimately because the solution $\p_0[\vek H]$ then minimizes the Hamiltonian density~\eqref{eq:effham} (with $m_\pi=0$) pointwise. For nonconservative magnetic fields, on the other hand, both the decomposition~\eqref{eq:helmholtz} and the ground state $\p_0[\vek H]$ depend sensitively on the choice of the domain $V$.


\subsection{Analytic solution from separation of variables}
\label{subsec:nonuniformBzeroMseparation}

For the sake of illustration, and to collect some benchmark results for testing our numerical code discussed below, we will now work out analytically the CSL ground state for the special case of a two-dimensional magnetic field, pointing everywhere in the same direction. Denoting the two relevant Cartesian coordinates as $x$ and $z$, we will orient the coordinate axes so that the magnetic field vector in the $xz$-plane can be written as
\begin{equation}
\vek H(x,z)=(0,H(x)).
\label{eq:2d_straightH}
\end{equation}
This generalizes our previous discussion of the CSL in a uniform magnetic field, chosen to point along the $z$-axis, whereby the magnitude of the field can now depend on $x$.

The choice of the magnetic field suggests that we search for the solution of the Laplace equation with the boundary condition~\eqref{eq:naturalBC} using separation of variables. To streamline the analysis, we therefore further assume that the domain $V$ takes the shape of a rectangle,
\begin{equation}
-\frac{L_x}2\leq x\leq+\frac{L_x}2,\qquad
-\frac{L_z}2\leq z\leq+\frac{L_z}2.
\label{eq:rectangle}
\end{equation}
The derivation of the solution is then a simple exercise in mathematical techniques familiar, for instance, from electrostatics. We therefore merely quote the final result,
\begin{align}
\label{eq:solseparation}
&\p_0(x,z)=\frac{z}{L_x}\int_{-{L_x}/2}^{+{L_x}/2}\dd t\,H(t)\\
\notag
&+\sum_{\text{even }n>0}\frac{2\int_{-{L_x}/2}^{+{L_x}/2}\dd t\,H(t)\cos\frac{\pi nt}{L_x}}{\pi n\cosh\frac{\pi nL_z}{2L_x}}\cos\frac{\pi nx}{L_x}\sinh\frac{\pi nz}{L_x}\\
\notag
&+\sum_{\text{odd }n>0}\frac{2\int_{-{L_x}/2}^{+{L_x}/2}\dd t\,H(t)\sin\frac{\pi nt}{L_x}}{\pi n\cosh\frac{\pi nL_z}{2L_x}}\sin\frac{\pi nx}{L_x}\sinh\frac{\pi nz}{L_x}.
\end{align}

The special case of particular interest we will analyze in greater detail below is that of a magnetic domain wall, $H_\text{DW}(x)=H_0\operatorname{sgn}x$. For this choice of magnetic field, as well as for any other $H(x)$ that is an odd function of $x$, the first two integrals in Eq.~\eqref{eq:solseparation} vanish. The last integral is nonzero and easy to evaluate explicitly, leading to the domain wall solution, depending on a sole scaling parameter $H_0$,
\begin{equation}
\begin{split}
\p_{\text{DW}}(x,z)=\sum_{\text{odd }n>0}&\frac{4H_0L_x}{\pi^2n^2\cosh\frac{\pi nL_z}{2L_x}}\\
&\times\sin\frac{\pi nx}{L_x}\sinh\frac{\pi nz}{L_x}.
\end{split}
\label{eq:DWsolution}
\end{equation}
In Appendix~\ref{app:domainwall}, we calculate in detail the average energy density of this solution as a function of the aspect ratio $L_z/L_x$ of the rectangular domain. This clarifies among others to what extent the CSL state in presence of a magnetic domain wall differs from the solution in a uniform magnetic field.

As an aside, note that the expression~\eqref{eq:DWsolution} can be further simplified if we picture the magnetic domain wall as separating two infinitely extended domains. Namely, keeping $L_z$ finite but taking the limit $L_x\to\infty$, the sum over $n$ can be replaced with an integral, leading to
\begin{equation}
\p_{\text{DW}}(x,z)\xrightarrow{L_x\to\infty}\frac{H_0}{\pi}\int_{-\infty}^{+\infty}\dd k\,\frac{\sin kx\sinh kz}{k^2\cosh\frac{kL_z}2},
\end{equation}
where $k$ is identified with $\pi n/L_x$ in the previous discrete sum.


\subsection{Digression: nonuniform baryon chemical potential}
\label{subsec:nonuniformBzeroMdigression}

An interesting modification of the basic setup for massless pions that preserves the linearity of the EoM is to consider a system where the baryon chemical potential $\m$ is a nontrivial but fixed function of spatial coordinates. This might be relevant for instance for inhomogeneous systems in local thermodynamic equilibrium. The bulk EoM for the CSL ground state then changes to
\begin{equation}
\grad^2\p_0=\divg\vek H=\frac{\vek B\cdot\grad\m}{4\pi^2f_\pi^2},
\end{equation}
which is a Poisson equation for $\p_0[\vek H]$ with the same Neumann boundary condition~\eqref{eq:naturalBC}.

In spite of the different EoM, much of the reasoning in Sec.~\ref{subsec:nonuniformBzeroMgeneral} can be directly applied to the present more general case. The gradient--solenoid decomposition~\eqref{eq:helmholtz} still holds, we just have to keep in mind that the rescaled magnetic field $\vek H$ as defined by Eq.~\eqref{eq:defH} is no longer solenoidal. Accordingly, the scalar potential $\psi[\vek H]$ satisfies the Poisson equation $\grad^2\psi=\divg\vek H$ in $V$ rather than the Laplace equation. Importantly, $\vek A[\vek H]$ still has vanishing divergence, and vanishing normal component on the boundary. We therefore conclude that the unique CSL ground state $\p_0[\vek H]$  still coincides with the scalar potential $\psi[\vek H]$ of $\vek H$.

In fact, it is easy to check that the derivation of the bounds~\eqref{eq:energybound1} and~\eqref{eq:energybound2} does not use anywhere the assumption that $\vek H$ is solenoidal. We can therefore repeat all our previous conclusions concerning the characterization of the CSL solution verbatim, as long as we phrase them in terms of the vector field $\vek H$ rather than $\vek B$, and keep in mind that the coordinate dependence of $\vek H$ is controlled by the convolution of the coordinate dependence of the physical magnetic field $\vek B$ and the baryon chemical potential $\m$. In particular, it is still true that the CSL ground state is energetically favored over the trivial QCD vacuum unless $\vek B$ is tangential to the boundary everywhere on $\de V$, a property that is unaffected by the coordinate dependence of the chemical potential. On the other hand, reaching the lower bound given by the right-hand side of Eq.~\eqref{eq:energybound2} requires that $\vek H$ rather than $\vek B$ be conservative, which is equivalent to the condition
\begin{equation}
\grad\m\times\vek B+\m\grad\times\vek B=\vek0.
\end{equation}


\section{Nonuniform magnetic fields for massive pions}
\label{sec:nonuniformBnonzeroM}

We now move to the physically most interesting, but also technically most challenging, case where the pion mass is not neglected and the external magnetic field is nonuniform. Here we will have to largely rely on direct numerical minimization of the energy functional. However, there is still a little we can say on a general ground without detailed computations.

First, note that the contribution of the pion mass to the Hamiltonian density~\eqref{eq:effham} is non-negative. Thus, the total energy of a given pion field configuration in a given magnetic field will always be greater than or equal to the energy of the same configuration in the same magnetic field in the massless limit. In particular, for magnetic fields that are tangential to the boundary everywhere on $\de V$, the energy is by Eq.~\eqref{eq:energybound1} bound to be non-negative.

Second, guided by our past success, we attempt to use the scalar potential $\psi[\vek H]$ of the magnetic field, defined by the decomposition~\eqref{eq:helmholtz}, as a trial variational ground state. We find that the energy of this specific scalar field is
\begin{equation}
E[\psi]=f_\pi^2\int_V\left[-\frac12(\grad\psi)^2+m_\pi^2(1-\cos\psi)\right].
\label{eq:massivebound}
\end{equation}
Here we have a good chance that the integrand in Eq.~\eqref{eq:massivebound} is negative everywhere. The energy of $\psi[\vek H]$ would then be negative, and by the variational principle, the energy of the actual ground state would necessarily lie at or below the same value. Yet, this argument does not look tremendously useful, since $\psi[\vek H]$ in general depends on both the magnetic field and the choice of the domain $V$. The exception to the rule are conservative magnetic fields, for which $\grad\psi[\vek H]=\vek H$. The integrand in Eq.~\eqref{eq:massivebound} will then be negative everywhere if $|\vek H|>2m_\pi$, which is easily converted using Eq.~\eqref{eq:defH} to a bound for $|\vek B|$.

To summarize, we can make two exact statements on the nature of the ground state:
\begin{itemize}
\item The trivial QCD ground state $\p_0=0$ prevails for magnetic fields that are tangential to the boundary everywhere on $\de V$. For such fields, there is no pion field configuration with negative total energy.
\item The ground state is of the CSL type for conservative magnetic fields whose magnitude is everywhere bounded from below by
\begin{equation}
|\vek B|>\frac{8\pi^2f_\pi^2m_\pi}{\m}=\frac\pi2B_\text{CSL},
\label{eq:BCSL:massivepi}
\end{equation}
regardless of the choice of the domain $V$. To appreciate the scale of magnetic fields in question, let us take the physical values of $f_\pi\approx92~\text{MeV}$ and $m_\pi\approx140~\text{MeV}$, and estimate $\mu=1~\text{GeV}$. Then, Eq.~\eqref{eq:BCSL} gives $B_\text{CSL}\approx0.06~\text{GeV}^{2}$, which is of the order of $10^{19}~\text{G}$. The bound~\eqref{eq:BCSL:massivepi} is then satisfied for $|\vek B|\gtrsim0.09~\text{GeV}^{2}$.
\end{itemize}
Remarkably, the threshold for magnetic fields sufficient to guarantee the appearance of a CSL-type state is fairly close to the similar bound~\eqref{eq:BCSL} for uniform magnetic fields. Note that this threshold is not in contradiction to the phase diagram in Fig.~\ref{fig:PD1d}. Namely, the curve therein is defined by a comparison of the energies of two actual solutions of the EoM. On the other hand, here we are making a case for field configurations with energy lower than that of the trivial vacuum, $\p_0=0$.


\subsection{Setting the stage: one-dimensional numerics}
\label{subsec:nonuniformBnonzeroMbenchmark}

Finding the ground state for a generic choice of the inhomogeneous magnetic field, and massive pions, requires direct numerical minimization of the energy functional based on the Hamiltonian density~\eqref{eq:effham}. For that purpose, we have developed a numerical code using \textsc{Wolfram Mathematica} (version $15.0$). Here we will first describe in some detail a one-dimensional version of the code, which we could test against the semi-analytic results of Sec.~\ref{sec:uniformB}. Further below, we will discuss the scaled-up two-dimensional version of the code along with sample numerical results for the ground state in inhomogeneous two-dimensional magnetic fields.

In order to compute the total energy of the pion field configuration, we have discretized the integration domain into a finite lattice with constant spacing. The derivative of the field was approximated using the symmetric difference quotient, except for the boundary points, where we used the simple one-sided difference quotient. The Hamiltonian density was subsequently integrated over the grid using the trapezoidal rule. The latter is considerably more accurate than the naive rectangular rule, and does not yet suffer from numerical artifacts that plague the application of higher-order integrators to inhomogeneous states of matter. Both of the above steps were implemented from scratch. The last step is to minimize the total energy with respect to the values of the pion field on the grid, which we implemented using \textsc{Mathematica}'s built-in global optimization routines.

\begin{figure}[t]
\centering
\includegraphics[width=\columnwidth]{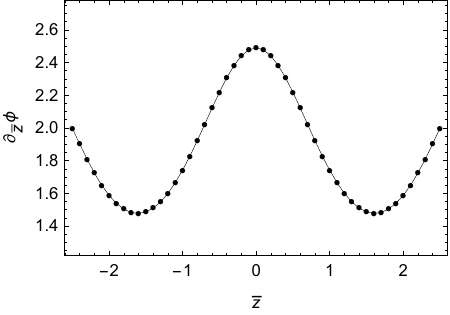}
\caption{Sample numerical solution for the CSL ground state in one dimension (uniform magnetic field), here for $\bar L=5$ and $\bar H=2$. In order to avoid the ambiguity with respect to shifts of the field $\p$ by multiples of $2\pi$, we display the gradient (derivative) of the solution, $\de_{\bar z}\p$. The data points indicate the numerical solution on a lattice with 50 unit cells, whereas the solid line beneath them shows the analytic solution obtained using the approach of Sec.~\ref{subsec:uniformBfiniteV}.}
\label{fig:1d_sample}
\end{figure}

As a simple consistency check, we compared the fully numerical computation of the ground state for a uniform magnetic field in one dimension with the semi-analytic results of Sec.~\ref{subsec:uniformBfiniteV} for a range of choices of the dimensionless system size $\bar L=m_\pi L$ and the dimensionless magnetic field $\bar H=H/m_\pi$. An example is shown in Fig.~\ref{fig:1d_sample}. This verifies that, at least in one dimension, the code finds the actual absolute minimum of the energy functional, not just one of the many local minima. Also, the profile of the ground state as represented by the gradient of the pion field is reproduced with high accuracy.

To further verify that the successful reproduction of the ground state as shown in Fig.~\ref{fig:1d_sample} is not a mere lucky accident, we have sampled a part of the parameter space of the dimensionless variables $\bar L$ and $\bar H$. Instead of checking visually the profile of the ground state case by case, we have implemented a simple numerical indicator, measuring the average (absolute value of the) difference of the gradient of the numerically found minimum and that of the analytically known ground state. The rationale behind this test is that values of the indicator betraying a misidentification of the ground state are typically orders of magnitude larger than values arising from mere numerical noise. This makes it easy to choose and implement a cutoff value of the indicator that discriminates between the two possibilities. We sampled the part of the parameter space defined by $0<\bar L\leq10$ and $0<\bar H\leq5$ with a $50\times50$ grid. Our numerical code has identified autonomously the absolute minimum of the energy functional for all but five points of the 2500 grid points. The failed attempts all occurred near curves in the parameter space where the energy functional has degenerate local minima.


\subsection{Two-dimensional numerics}
\label{subsec:nonuniformBnonzeroMsample}

The same strategy was deployed to find the ground state on the two-dimensional rectangular domain~\eqref{eq:rectangle} with a generic inhomogeneous magnetic field. We discretized the domain into an $n_x\times n_z$ grid, approximated the gradient of the pion field by central difference quotients, and integrated the Hamiltonian density using the two-dimensional version of the trapezoidal rule, which gives the discretized total energy as a function of the grid variables. We stress that there is no need to impose any boundary condition by hand; the discretized variational problem automatically implements the natural boundary condition~\eqref{eq:naturalBC}.

Importantly, increasing the number of dimensions scales up the number of grid points and thus of variables to optimize for. This poses a serious challenge to finding the ground state on sufficiently fine grids. To ameliorate the issue, we combined \textsc{Mathematica}'s global optimizer \textsc{NMinimize} with the much faster but local optimizer \textsc{FindMinimum}. Typically, we would use the latter on a fine grid to produce the profile of the ground state, and the former on a coarser grid to cross-check that the ground state has been identified qualitatively correctly.

As is clear from the expression~\eqref{eq:effham} for the Hamiltonian density, the ground state depends on the pion mass $m_\pi$ and the length scales $L_x,L_z$, plus whatever other variables parameterize the chosen model for the magnetic field. Since the energy functional can be rescaled arbitrarily without affecting its stationary states, the number of relevant parameters can be reduced by considering only dimensionless combinations of all the input variables. In contrast to what we have largely done until now, we will not set the units by rescaling with $m_\pi$ in order to keep open the possibility to interpolate between the cases of massive and massless pions for the sake of comparison. Instead, we will take the liberty to assign the different input variables numerical values with respect to an arbitrarily chosen scale, keeping in mind that only dimensionless combinations such as $L_z/L_x$ or $m_\pi L_x$ are relevant.

As a basic consistency check of the setup, we applied our two-dimensional code to the case of vanishing pion mass and a uniform magnetic field, corresponding to $H(x)=H_0$ in Eq.~\eqref{eq:2d_straightH}. The numerically found average energy density of the ground state agrees with the analytic result $-f_\pi^2H_0^2/2\equiv\Ea_0$ with high precision for all choices of the rectangular domain~\eqref{eq:rectangle} and of the $n_x\times n_z$ grid considered. Moreover, the ground state produced by the numerical code agrees perfectly with the analytic result~\eqref{eq:vacuniformchiral}. This stems from the fact that the exact pion field profile is linear in $z$, and is therefore represented on the grid without any discretization error.

\begin{figure}[t]
\centering
\includegraphics[width=\columnwidth]{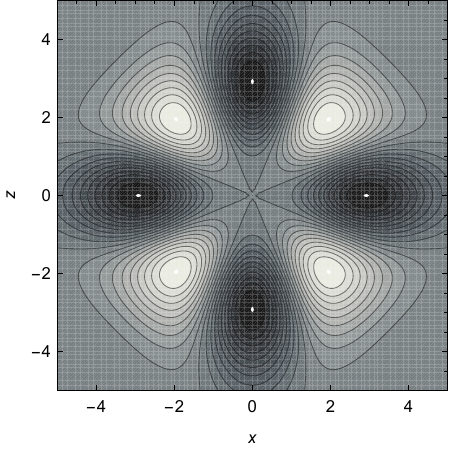}
\caption{Baryon number density $\vek H\cdot\grad\p$ (in arbitrary units) of the CSL ground state in a Gaussian magnetic field of the type~\eqref{eq:GaussH} with $\psi_0=4$ and $R=2$, on a $L_x\times L_z=10\times10$ domain in the chiral limit. The computation was carried out by a direct minimization of the energy functional discretized on a $51\times51$ grid. The displayed contours were obtained by interpolation of the grid data for the pion field. The grayscale runs from dark at the smallest to light at the largest value of $\vek H\cdot\grad\p$. Thus, the four bright regions on the diagonals represent peaks of positive baryon density and the darker regions along the axes peaks of negative baryon density.}
\label{fig:Gaussian_field}
\end{figure}

To show one example of a CSL solution in a nontrivial magnetic field before we move on, we have computed the ground state for a two-dimensional Gaussian magnetic field, $\vek{H}=(-\de_z\psi,\de_x\psi)$ with the scalar potential 
\begin{equation}
\psi(x,z)=\psi_0\exp\left(-\frac{x^2+z^2}{R^2}\right).
\label{eq:GaussH}
\end{equation}
For concreteness, we chose $\psi_0=4$ and $R=2$ (in arbitrary units), and minimized the energy functional on a rectangular domain with $L_x=L_z=10$. In Fig.~\ref{fig:Gaussian_field}, we show the ground state baryon number density, $n_\text{B}=\vek B\cdot\grad\p/(4\pi^2)$, rescaled by replacing the physical magnetic field $\vek B$ with $\vek H$ as defined by Eq.~\eqref{eq:defH}.


\subsection{Case study: magnetic domain wall}
\label{subsec:nonuniformBnonzeroMdomainwall}

By way of a case study, we will now discuss in some detail the properties of the CSL ground state in a domain-wall-like magnetic field. Here we already have some benchmark results from Sec.~\ref{subsec:nonuniformBzeroMseparation}, which we will further generalize by considering magnetic fields whose profile resembles a domain wall with a nonzero width. The motivation for focusing on the magnetic domain wall comes from first-principle lattice simulations of QCD. Namely, in lattice simulations, the spacetime is effectively compactified to a torus. A uniform magnetic field has a nonzero flux through the torus, which leads to a quantization condition on the allowed values of the magnetic field. To avoid this problem, it has been proposed to use a magnetic field that is constant on a half of the lattice, and opposite on the other half~\cite{Levkova2014}. This makes the flux vanish, and thus allows to tune the magnetic field strength at will. 

The class of domain-wall-type magnetic fields we will consider is defined by Eq.~\eqref{eq:2d_straightH} along with
\begin{equation}
H_\xi(x)=H_0\tanh\frac{x}{\xi},
\label{eq:tanhwall}
\end{equation}
where the positive parameter $\xi$ is interpreted as the width of the domain wall. Tuning the value of the width interpolates between the extreme cases of a sharp domain wall with $\xi\to0$ and $H_\text{DW}(x)=H_0\operatorname{sgn}x$ and a magnetic field with $\xi\gtrsim L_x$ that is effectively linear in $x$ throughout the rectangular domain~\eqref{eq:rectangle}.

\begin{figure}[t]
\centering
\includegraphics[width=\columnwidth]{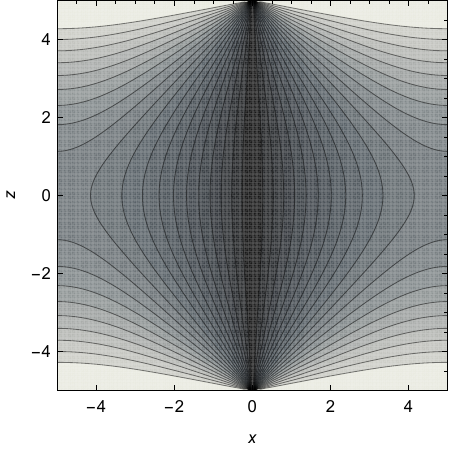}
\caption{Rescaled baryon number density $\vek H\cdot\grad\p$ in the same arbitrary units as in Fig.~\ref{fig:Gaussian_field}, for the CSL ground state in a domain wall magnetic field~\eqref{eq:tanhwall} with $H_{0}=5$ and $\xi\to 0$, in the chiral limit, on the domain $L_x\times L_z=10\times 10$. The baryon number density vanishes along the domain wall localized at $x = 0$ where the field changes sign. The computation was carried out by a direct minimization of the energy functional discretized on a $101 \times 101$ grid.}
\label{fig:Baryon_density_DW}
\end{figure}

For the sake of simplicity, we will mostly consider square-shaped domains where $L_x=L_z\equiv L$. This reduces the relevant parameter space to three dimensionless combinations of $m_\pi$, $H_0$, $\xi$ and $L$.  Unless explicitly specified otherwise, the numerical results reported below were obtained by fixing $H_0=5$ in arbitrary units on a $31\times31$ grid.

To begin with, we validated our two-dimensional code against the analytic results of Sec.~\ref{subsec:nonuniformBzeroMseparation}. Namely, we used the code to address the case of a sharp domain wall ($\xi\to0$) in the massless limit. Here we know from Appendix~\ref{app:domainwall} that for a square-shaped domain, the CSL ground state should have average energy density $\langle\Ea_\text{DW}\rangle=\Ea_0/2$. Our numerical code gives a ground state with an average energy density of approximately $0.499\Ea_0$, independently of the actually chosen value of $L_x=L_z\equiv L$ over the entire range of $L$ considered. The numerical error of about $0.2\%$ is consistent with discretization on a $31\times31$ grid and decreases upon further refinement of the grid. To visualize the CSL profile in the sharp domain wall magnetic field, we show in Fig.~\ref{fig:Baryon_density_DW} the baryon number density for a specific choice of the input parameters $H_0$ and $L$.

\begin{figure}[t]
\centering
\includegraphics[width=\columnwidth]{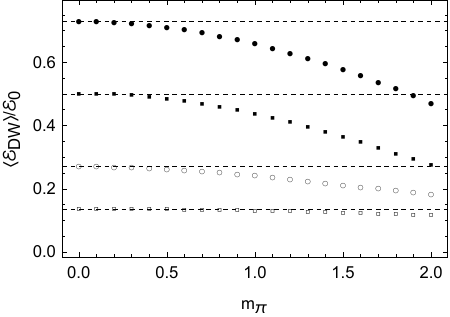}
\caption{Normalized average energy density of the CSL ground state for a sharp magnetic domain wall ($\xi\to0$) as a function of the pion mass and for several values of the aspect ratio $\a\equiv L_z/L_x$. All the calculations were done with $H_0=5$ and $L_z=4$; the aspect ratio was tuned by varying $L_x$ accordingly. The minimization of the energy functional was performed on a $31\times31$ grid. The filled circles, filled squares, open circles and open squares correspond respectively to $\a=0.5$, $\a=1$, $\a=2$ and $\a=4$. The $m_\pi\to0$ intercepts reproduce the analytic result of Appendix~\ref{app:domainwall}, indicated by the dashed horizontal lines.}
\label{fig:energy_mass}
\end{figure}

To inspect the effect of the pion mass, we have computed the average energy density of the CSL solution for the sharp magnetic domain wall ($\xi\to0$) as a function of $m_\pi$, for several choices of the rectangular domain~\eqref{eq:rectangle} with different aspect ratios $\a\equiv L_z/L_x$. This setup is designed to directly test the deviations from the analytical expression~\eqref{eq:DWenergy_final} due to nonzero pion mass. The numerical results are shown in Fig.~\ref{fig:energy_mass}. On the one hand, they check the consistency of our numerical code by reproducing the predictions of Appendix~\ref{app:domainwall} in the chiral limit. On the other hand, they give a quantitative measure for the effects of tuning the pion mass. 

\begin{figure}[t]
\centering
\includegraphics[width=\columnwidth]{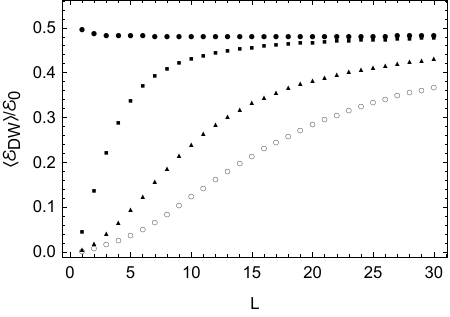}
\caption{Normalized average energy density of the CSL ground state on a square domain ($L_x=L_z=L$) of varying size for domain wall magnetic fields~\eqref{eq:tanhwall} with $H_0=5$ and several different values of the width $\xi$. The filled circles, filled squares, filled triangles and open circles correspond respectively to $\xi\to 0$, $\xi=1$, $\xi=3$ and $\xi=5$. The computation was carried out by minimization of the energy functional on a $31\times 31$ grid. The pion mass was kept fixed to $m_\pi=0.5$ in arbitrary units.}
\label{fig:energy_length}
\end{figure}

In order to get another angle on the parameter space associated with domain wall magnetic fields, we have computed the energy density of the CSL ground state on a square-shaped domain as a function of the system size $L$ for several values of the domain wall width $\xi$. The results are shown in Fig.~\ref{fig:energy_length}. We find that, unsurprisingly, for $L\lesssim\xi$, the condensation energy tends to zero since the magnetic field is very weak throughout the domain on which the system lives. The pion condensation sets in for $L\gg\xi$, as one can clearly see by comparing the curves corresponding to different values of $\xi$.

We conclude the discussion with a word of caution. Due to the discretization of the optimization problem on a spatial grid, one should be careful and not vary the input parameters mindlessly while keeping the grid size $n_x\times n_z$ fixed. Ideally, one should take the continuum limit by increasing the number of grid points and extracting the asymptotic behavior of the results in the limit $n_{x,z}\to\infty$, but this is of course challenging as it involves a rapid growth of the number of variables one needs to optimize for. In the numerical computations reported in this paper, we mostly chose to work with a fixed number of grid points. To estimate whether such an approach is meaningful, note that the dimensions of an individual cell of the lattice are $L_x/(n_x-1)\times L_z/(n_z-1)$. On the other hand, our EFT possesses an intrinsic scale given by the Compton wavelength of the pion, $1/m_\pi$. We can expect discretization artifacts to be reasonably small provided the size of the lattice cell is (much) smaller than the Compton wavelength, that is, for $n_x\gtrsim m_\pi L_x$ and $n_z\gtrsim m_\pi L_z$. As an illustration, a uniform field with $H_{0} = 5$ and $m_{\pi} = 0.5$ on a domain of size $L = 20$ gives an average energy density of $-12.36$ on a $ 31 \times 31$ grid, against the exact value $-12.25$. The discretization therefore leads to the error of only about $1\%$, in accord with the fact that the condition $n_{x,z}\gtrsim m_\pi L_{x,z}$ is well satisfied.  


\section{Summary and outlook}
\label{sec:summary}

In this paper, we have initiated the exploration of the effects of magnetic field inhomogeneity on the CSL phase of QCD. The main result is that even inhomogeneous magnetic fields may support a CSL-type ground state, depending on the precise choice of the magnetic field and of the domain in which the system lives. Remarkably, for the special class of conservative magnetic fields and in the limit of vanishing pion mass, the inhomogeneity does not present any obstacle whatsoever to the formation of the CSL state. Namely, the gradient of the neutral pion field is locally equal, up to a constant factor, to the magnetic field just like in the previously studied special case of uniform fields. Even for nonzero pion mass, conservative magnetic fields favor a CSL ground state, at least above certain threshold for the magnitude of the field as given by Eq.~\eqref{eq:BCSL:massivepi}.

We have analyzed specifically the CSL configuration for magnetic fields of the domain wall type~\eqref{eq:tanhwall}, relevant for first-principle lattice simulations of QCD. Apart from their relevance for lattice simulations, such fields are also interesting conceptually, for they constitute a prototype of fields whose inhomogeneity is localized in space. This allows for a particularly clean comparison to the well-known uniform-field limit. We found that also domain wall magnetic fields, though not being conservative, support a CSL ground state. In the chiral limit, the deviation of the corresponding CSL solution from that for uniform magnetic fields can be characterized semi-analytically by the spatial average of condensation energy density of the neutral pion field, as shown in Appendix~\ref{app:domainwall}.

There are at least two natural avenues for follow-up research in this direction. On the one hand, the ground state itself is but one outstanding property of the CSL phase. Equally interesting is the excitation spectrum, which affects the transport properties of the CSL state and its thermodynamical properties at nonzero temperature. In this regard, it would be interesting, though presumably highly nontrivial, to investigate more closely what kind of excitations the CSL state in nonuniform magnetic fields can host. On the other hand, it would be desirable to repeat the analysis presented here for phenomenologically more realistic magnetic fields, which would give us a better estimate for the practical importance of the effects of magnetic field inhomogeneity on the CSL phase. We leave both of the above equations to the future work.


\backmatter
\section*{Acknowledgements}

T.B.~would like to thank G.~Endr\H{o}di for pointing out the relevance of domain wall magnetic fields for lattice simulations of QCD. R.R.~would like to thank A.~Andis, T.~Kar and P.~Manoprakash for their help with \textsc{Mathematica}.


\section*{Statements and declarations}

The authors have no competing interests to declare that are relevant to the content of this article.


\begin{appendices}

\section{Energy of the domain wall in the massless limit}
\label{app:domainwall}

In this appendix, we will take a closer look at the properties of the CSL solution~\eqref{eq:DWsolution} for the magnetic domain wall, $\vek H(x,z)=(0,H_0\operatorname{sgn}x)$, on the rectangular domain~\eqref{eq:rectangle}. Our goal is to compute the energy of this solution and contrast it to the energy of the solution in the uniform field $\vek H(x,z)=(0,H_0)$. This will give us insight in how much the domain wall localized at $x=0$ affects the ground state far away from it. For the record, we note that the energy of the solution~\eqref{eq:vacuniformchiral} in the uniform field is
\begin{equation}
E[\p_0]=-\frac{f_\pi^2}{2}H_0^2L_xL_z\equiv\Ea_0L_xL_z.
\label{eq:Evacuniform}
\end{equation}

For the CSL ground state $\p_0$ with a generic choice of the magnetic field, the energy is obtained by replacing $\psi[\vek H]$ by $\p_0$ on the right-hand side of Eq.~\eqref{eq:energybound1}. Inserting the solution~\eqref{eq:DWsolution}, using the orthogonality of the Fourier basis functions, and dividing by the area $L_xL_z$ of the integration domain, we readily obtain an expression for the average energy density,
\begin{equation}
\langle\Ea_\text{DW}\rangle=16\Ea_0\frac{L_x}{L_z}\sum_{\text{odd }n>0}\frac{1}{(\pi n)^3}\tanh\frac{\pi nL_z}{2L_x}.
\label{eq:DWaverageE}
\end{equation}
This only depends on the aspect ratio of the rectangular domain, $L_z/L_x\equiv\a$, not on its actual size. It is straightforward to check that for a square domain with $\a=1$, $\langle\Ea_\text{DW}\rangle=\Ea_0/2$, whereas for an effectively one-dimensional domain with $\a\to0$, we find $\langle\Ea_\text{DW}\rangle\to\Ea_0$. The latter limit confirms that the effect of the magnetic domain wall on the CSL solution is localized. To understand more precisely how much of the condensation energy stored in the pion field can be associated with the domain wall, we need to inspect the behavior of the difference $\langle\Ea_\text{DW}\rangle-\Ea_0$ for small $\a$.

To that end, we will derive an asymptotic expansion of the ratio $\langle\Ea_\text{DW}\rangle/\Ea_0$ in $\a$. There is an elementary derivation that uses as the sole input the following summation identity, known from thermal field theory where it appears as a fermionic Matsubara sum,
\begin{equation}
\sum_{\text{odd }k>0}\frac1{k^2+t^2}=\frac\pi{4t}\tanh\frac{\pi t}2,
\label{eq:Matsubara}
\end{equation}
with arbitrary real $t$. Using this identity, the average energy density given by Eq.~\eqref{eq:DWaverageE} can be rewritten as
\begin{align}
\frac{\langle\Ea_\text{DW}\rangle}{\Ea_0}&=\frac{64\a^2}{\pi^4}\sum_{\substack{\text{odd }n>0\\\text{odd }k>0}}\frac1{(n\a)^2[k^2+(n\a)^2]}\\
\notag
&=\frac{64\a^2}{\pi^4}\sum_{\substack{\text{odd }n>0\\\text{odd }k>0}}\frac1{k^2}\biggl[\frac1{(n\a)^2}-\frac1{k^2+(n\a)^2}\biggr].
\end{align}
In the first term on the right-hand side, both sums can be carried out trivially. In the second term, we use Eq.~\eqref{eq:Matsubara} to work out the sum over $n$. In the result, we replace $\tanh[\pi k/(2\a)]$ with $1-2/(e^{\pi k/\a}+1)$. The part without the exponential can then be summed over $k$ analytically. Putting all the pieces together, we arrive at an exact representation of the normalized average energy density of the CSL solution for a magnetic domain wall,
\begin{equation}
\frac{\langle\Ea_\text{DW}\rangle}{\Ea_0}=1-\frac{14\a\zeta(3)}{\pi^3}+\frac{32\a}{\pi^3}\sum_{\text{odd }k>0}\frac{1}{k^3}\frac{1}{e^{\pi k/\a}+1}.
\label{eq:DWenergy_final}
\end{equation}

\begin{figure}[t]
\centering
\includegraphics[width=\columnwidth]{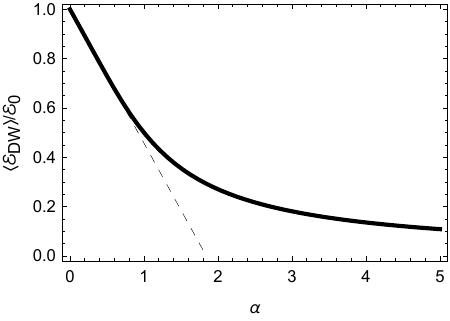}
\caption{Normalized average energy density of the CSL ground state for the magnetic domain wall, given by the right-hand side of Eq.~\eqref{eq:DWenergy_final}, as a function of the aspect ratio $\a\equiv L_z/L_x$. The dashed line indicates the first two terms of the asymptotic expansion, that is the constant and linear term in Eq.~\eqref{eq:DWenergy_final}.}
\label{fig:DW}
\end{figure}

The first line of Eq.~\eqref{eq:DWenergy_final} constitutes the entire power expansion of the average energy density in $\a$; all the subleading corrections are nonanalytic and exponentially suppressed for small $\a$. The difference between the average energy densities of the CSL states for uniform and domain wall magnetic fields is thus dominated by the term linear in $\a$, which translates to a difference in the total (integral) energies that scales as $L_x^0L_z^2$. This corresponds physically to an energy excess that is localized on the domain wall, as expected. Between $\a=0$ and $\a=1$, the function~\eqref{eq:DWenergy_final} is nearly linear, showing that numerically, the nonanalytic subleading corrections play a minor role; see Fig.~\ref{fig:DW}. In the (somewhat less physical) limit of $\a\to\infty$, the asymptotic behavior of the average energy density of the CSL solution for the magnetic domain wall is most easily extracted by setting the $\tanh$ in Eq.~\eqref{eq:DWaverageE} to one, which leads to
\begin{equation}
\frac{\langle\Ea_\text{DW}\rangle}{\Ea_0}\xrightarrow{\a\to\infty}\frac{14\zeta(3)}{\pi^3\a}.
\end{equation}
This translates to a total (integral) energy that scales as $L_x^2L_z^0$.

\end{appendices}


\bibliography{references}

\end{document}